\documentstyle[epsfig]{aipproc}

\begin{document}
\title{Probing Anomalous Higgs Coupling through  $\mu^+\mu^- \to H \gamma$}

\tighten
\author{Ali Abbasabadi$^\dag$, David Bowser-Chao$^{\S,\P}$, \\ Duane A. Dicus$^*$ and 
Wayne W. Repko$^\ddag$}
\address{$^\dag$Department of Physical Sciences, Ferris State University, 
Big Rapids, Michigan 49307 \\
\smallskip
$^{\S}$Department of Physics,
University of Illinois at Chicago, Chicago, Illinois 60607; \\
\smallskip
$^*$Center for Particle Physics and Department of Physics, 
University of Texas, \\Austin, Texas 78712 \\
\smallskip
$^\ddag$Department of Physics and Astronomy, 
Michigan State University,\\ East Lansing, Michigan 48824\\
\smallskip
$^\P$Presented the talk.
}

\maketitle

\begin{abstract}
The process $\mu^+\mu^- \to H\gamma$, though small compared to the
resonance process, could be observable at proposed muon colliders, given
expected integrated luminosities. The apparently leading diagrams occur
at tree-level, and are proportional to the Higgs-muon coupling. We show,
however, that the one-loop contribution is comparable to the tree-level
amplitude. Furthermore, the one-loop diagrams, unlike those at
tree-level, could be greatly enhanced by  possible anomalous Higgs-top
quark or Higgs-gauge boson couplings. For a 500 GeV unpolarized muon
collider, the  total cross section for $H\gamma$ associated production
approaches 0.1 fb.
\end{abstract}

Recently, the possibility of using $\mu^+\mu^-$ colliders to investigate
the properties of  Higgs-bosons has received considerable attention
\cite{smass,gun}. There are significant advantages to studying
Higgs-bosons with this type of collider, particularly if the mass is
known from its discovery at, say, the LHC or NLC \cite{gun}. Under these
circumstances, the width and branching ratios can be studied at the
Higgs pole.

Resonance production, of course, is not the only channel for Higgs
production at such colliders. Production in association with a photon
has been considered in Ref.\cite{lt}, where the leading tree-level
diagrams (Fig. 1) involve radiation of the Higgs directly off the muon
line, and thus are proportional the Higgs-muon coupling, which in turn
(in the Standard Model) is proportional to the small but non-zero muon
mass.

On the other hand, the analogous process $e^+e^- \to H\gamma$ has been
studied\cite{ab-cdr}, and found to have a cross-section comparable to
the muon process, despite the relatively infinitesimal electron mass. In
this case, it turns out that the one-loop diagrams are orders of
magnitude larger than the tree-level diagrams, and in fact should not be
considered radiative corrections to the latter --- the former persist
even in the limit of zero electron mass. This process can be generalized
to the hadronic reaction $q\bar q \to H \gamma$\cite{qqhg}, and to the
rare decays $H\to f\bar f \gamma$, where the fermion $f$ may be
massless\cite{hdecay}. Turning back to the muon reaction, we shall see
below that the one-loop contribution is comparable to the tree-level
amplitude. Furthermore, the one-loop diagrams, unlike those at
tree-level, could be greatly enhanced by  possible anomalous Higgs-top
quark or Higgs-gauge boson couplings.

\begin{figure}[b!] 
\centerline{\epsfig{file=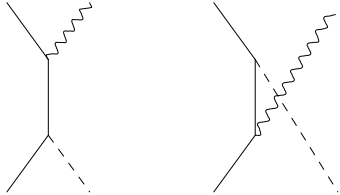,height=1.0in}}
\vspace{10pt}
\caption{Tree level diagrams for \protect$\mu^+\mu^-\protect\rightarrow
H\gamma\protect$ are shown.}
\label{fig1}
\end{figure}

For both the tree-level and loop-level diagrams discussed in detail
below, we label the momentum and helicity of the $\mu^-$ ($\mu^+$) in
the center of mass frame, respectively, by $p=(E,|{\bf p}| \hat z)$
($\bar p=(E,-|{\bf p}| \hat z)$) and $\lambda$ ($\bar \lambda$), the
photon  momentum and scattering angle by $k=(\omega ,\omega \hat k)$ and
$\theta$, and the photon polarization vector and helicity by $\epsilon$
and $\lambda_{\gamma} = \pm 1$. The tree level diagrams for
$\mu^+\mu^-\rightarrow H\gamma$ are shown in Fig. 1. The amplitudes,
labelled by muon and photon helicities, are found to be: 
\begin{eqnarray} \label{treeamp}
{\cal M}^{\rm tree}_{\lambda\bar{\lambda}\lambda_{\gamma}} &= &
-i\frac{egm_{\mu}}{\sqrt{\displaystyle 2}\,m_W}\left(\frac{1}{2p\!\cdot\!k} + 
\frac{1}{2\bar{p}\!\cdot\!k}\right) \nonumber \\
&&\cdot
\left\{
\begin{array}{lcl}
\sin\theta\left[\lambda_{\gamma}(2|{\bf p}|^2 - E\omega) + |{\bf
p}|\omega \right] && \lambda\bar{\lambda} = ++ \\
\sin\theta\left[\lambda_{\gamma}(2|{\bf p}|^2 - E\omega) - |{\bf
p}|\omega \right]&& \lambda\bar{\lambda} = -- \\ m_{\mu}\,\omega(1 +
\lambda_{\gamma}\cos\theta) & & \lambda\bar{\lambda} = +- \\
m_{\mu}\,\omega(1 - \lambda_{\gamma}\cos\theta) && \lambda\bar{\lambda}
= -+ \\
\end{array}
\right.\,,
\end{eqnarray}
explicitly showing the extra suppression of the helicity flip amplitudes
by a factor of $m_{\mu}$  relative to the non-flip amplitudes (where
$\lambda=+$ denotes a $\mu^-$ helicity of $+1/2$.)

The one-loop amplitudes for $\mu^+\mu^-\rightarrow H\gamma$ receive
contributions from pole diagrams involving virtual photon and $Z$
exchange and from various box diagrams containing muons, gauge bosons
and/or Goldstone bosons \cite{ab-cdr,ddhr,mumu}. There are also double
pole diagrams whose contribution vanishes. These are illustrated in Fig.
2. In the non-linear gauges we chose \cite{ab-cdr}, the full amplitude
consists of four separately gauge invariant terms: a photon pole, a $Z$
pole, $Z$ boxes and $W$ boxes. These amplitudes take the form:
\begin{eqnarray}\label{g}
{\cal M}^{\gamma,Z}_{\rm pole} & = &\, {\cal P}_{\gamma,Z}(s)
\left(
  \delta_{\mu\nu}k\!\cdot\!(p + \bar{p}) - k_{\mu}(p + \bar{p})_{\nu}
  \right) \;
\bar{v}(\bar{p})\gamma_{\mu}(v^{\cal P}_{\gamma,Z} 
	+ a^{\cal P}_{\gamma,Z}\gamma_5)u(p)
\epsilon_{\nu}^*\,, \\ [4pt]
{\cal M}_{\rm box}^{\gamma,Z} & = &
\, 
\left[
{\cal B}_{\gamma,Z}(s,t,u)
	\left(\delta_{\mu\nu}k\!\cdot\!p - k_{\mu}p_{\nu}\right)
	+ p \leftrightarrow \bar{p} 
\right]
	\;
  \bar{v}(\bar{p})\gamma_{\mu}(v^{\cal B}_{\gamma,Z}
  	 + a^{\cal B}_{\gamma,Z}\gamma_5)u(p)
\epsilon_{\nu}^*\;, 
\end{eqnarray}
where $s = -(p + \bar{p})^2,t = -(p - k)^2\,$ and $u = -(\bar{p} -
k)^2$. We have explicitly evaluated  the form factors (${\cal
P}_\gamma$, etc.) in terms of the scalar functions defined in the
appendices of our previous paper \cite{ab-cdr}; the results, though more
complicated than those of the tree-level diagrams, are provided in
Ref.\cite{mumu} in closed form. In this case, as should be expected, it
is the helicity {\em flip} contributions from the factors
$\bar{v}(\bar{p})\gamma_{\mu}u(p)$ and
$\bar{v}(\bar{p})\gamma_{\mu}\gamma_5 u(p)$ which survive in the
$m_{\mu}\rightarrow 0$ limit, as can be seen, for example, by explicitly
evaluating the spinor products above.

\begin{figure}[b!] 
\centerline{\epsfig{file=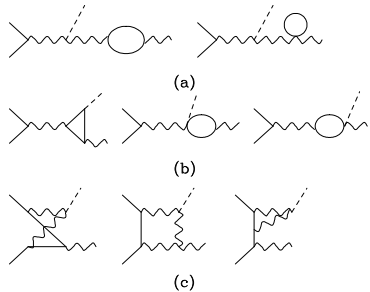,height=1.5in}}
\vspace{10pt}
\caption{Typical diagrams for the double pole (a), single pole (b) and
box (c) corrections are shown. An external solid line represents a muon,
a wavy line a gauge boson, a dashed line a Higgs boson and an internal
solid line a muon, gauge boson, Goldstone boson or ghost.}
\label{fig9}
\end{figure}

\begin{table}
\caption{The cross sections for the associated production of 100 GeV and
200 GeV Higgs bosons are shown together with the ratio of the signal to
the square root of the background ($S/\protect\sqrt{B}$) for several
$\mu^+\mu^-$ collider energies. A luminosity of 100 fb$^{-1}$ is
assumed.}
\begin{tabular}{ccccc}
$\sqrt{s}$ & $\sigma(m_H = 100\;{\rm GeV})$& $S/\sqrt{B}$&$\sigma(m_H = 200\;
{\rm GeV})$& $S/\sqrt{B}$ \\
\tableline
500 GeV  &$6.78\times 10^{-2}$\,fb &1.99
	& $8.76\times 10^{-2}$\,fb& 3.06 \\
1000 GeV &$2.46\times 10^{-2}$\,fb &1.50
	 & $3.87\times 10^{-2}$\,fb& 3.08 \\
2000 GeV &$8.76\times 10^{-3}$\,fb &1.06 
	& $1.04\times 10^{-2}$\,fb& 1.72 \\
4000 GeV &$1.54\times 10^{-3}$\,fb &0.36 
	& $2.17\times 10^{-3}$\,fb& 0.72
\end{tabular}
\end{table}

\begin{table}
\begin{center}
\caption{Cross sections for the background process $\mu^+\mu^-\protect
\rightarrow \gamma b\bar{b}$ are given for several cuts on the
$b\bar{b}$ invariant mass $m_{b\bar{b}}$. The last two columns are 5 GeV
bins indicating, respectively, the background associated with a Higgs
boson of mass 100 GeV or 200 GeV.}
\begin{tabular}{cccc}
$\sqrt{s}$ & $45\,{\rm GeV} < m_{b\bar{b}} < \sqrt{s}$ 
           & $97.5\;{\rm GeV} < 
m_{b\bar{b}} <102.5\,{\rm GeV}$ 
	       & $197.5\,{\rm GeV} < m_{b\bar{b}} < 202.5\; 
{\rm GeV}$ \\ 
\tableline 
500 GeV  & 11.1\,fb & $1.16\times 10^{-1}$\,fb  
	& $8.20\times 10^{-2}$\,fb \\
1000 GeV & 3.80\,fb & $2.69\times 10^{-2}$\,fb 
	 & $1.58\times 10^{-2}$\,fb \\
2000 GeV & 1.21\,fb & $6.83\times 10^{-3}$\,fb 
	 & $3.65\times 10^{-3}$\,fb \\
4000 GeV & 0.37\,fb & $1.81\times 10^{-3}$\,fb 
	 & $9.04\times 10^{-4}$\,fb 
\end{tabular}
\end{center}
\end{table}

The differential cross section $d\sigma(\mu^+\mu^-\rightarrow
H\gamma)/d\Omega _{\gamma}$ is given by 
\begin{equation}\label{dsig}
\frac{d\sigma(\mu^+\mu^-\rightarrow H\gamma)}{d\Omega_{\gamma}} =
\frac{1}{256\pi^2}\frac{s - m_H^2}{\beta s^2}\sum_{\rm spin}
|{\cal M}^{\rm tree} + {\cal M}^{\rm loop}|^2\,,
\end{equation}
with $\beta = \sqrt{1 - 4m_{\mu}^2/s}$. When integrating
Eq.\,(\ref{dsig}) to obtain the total cross section,  the   interference
terms should be suppressed, since ${\cal M}^{\rm tree}_{\pm\mp}$ and
${\cal M}^{\rm loop}_{\pm\pm}$ both contain an additional factor of
$m_{\mu}$. This conclusion can only be invalid if the angular
integration of the muon propagator factors $(1 \pm
\beta\cos\theta)^{-1}$ in Eqs.\,(\ref{treeamp}) produces inverse powers
of $m_{\mu}$. This is not the case. For the $++$ or $--$ interference
terms, the tree and loop amplitudes contain a factor of $\sin\theta$,
which ensures that the angular integral is well behaved in the
$\beta\rightarrow 1$ limit. The integral of $+-$ and $-+$ interference
terms can produce a factor of $\beta^{-1}$, but this also is finite as
$\beta\rightarrow 1$. As a consequence, we can simply add the tree
\cite{+-} and one-loop cross sections to obtain
$\sigma(\mu^+\mu^-\rightarrow H\gamma)$ for unpolarized muon beams.
Working to leading order in $m_\mu$, the polarized cross-sections are
obtained simply by omitting the tree-level or loop-level contributions
for helicity-flip or same-helicity beam polarization.

\begin{figure}[b!] 
\centerline{\epsfig{file=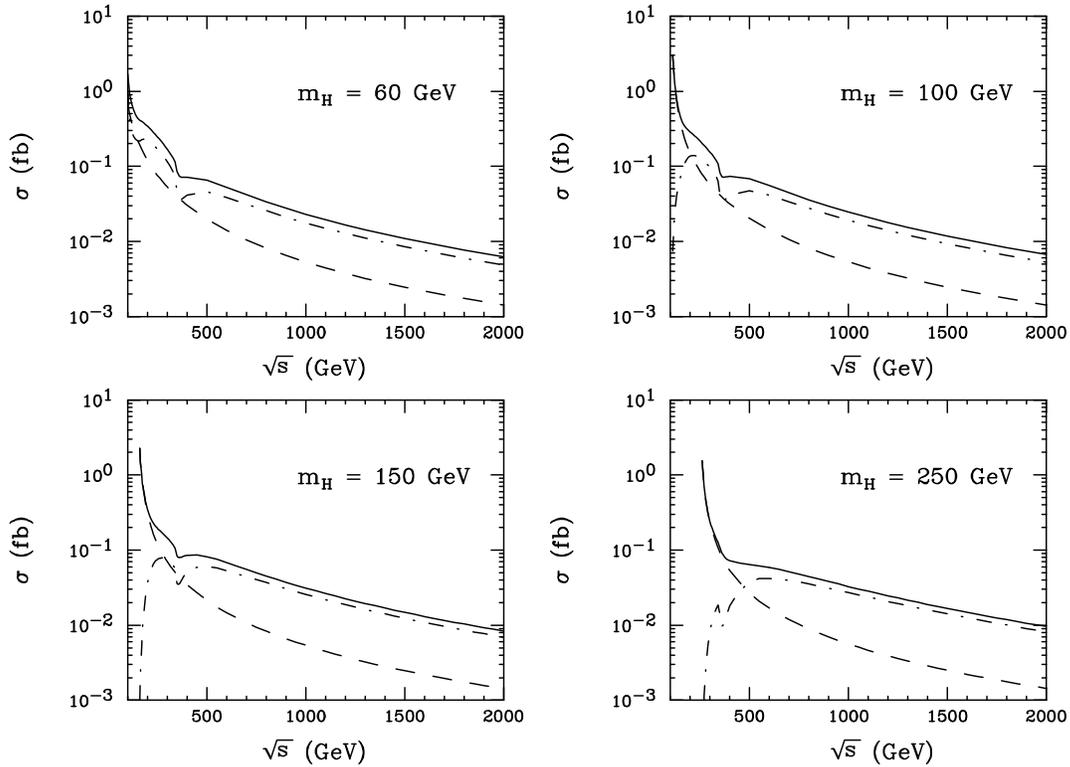,height=4.0in}}
\vspace{10pt}
\caption{The cross section for $\mu^+\mu^-\protect\rightarrow H\gamma$
resulting from the sum of the tree level and one-loop amplitudes is
given for several values of $m_H$ by the solid line. In each panel, the
dashed line is the tree level contribution and the dot-dashed line is
the one-loop contribution.}
\label{fig10}
\end{figure}

The result is illustrated in Fig. 3, where the tree, one-loop and total
cross sections are plotted for several values of $m_H$ as a function of
the collider energy. For collider energies $\sqrt{s}$ above about
$500$\,GeV, the one-loop contribution exceeds the tree contribution.
Note that Fig. 3 should not be taken literally at $\sqrt{s} \approx
m_H$. As $\omega\rightarrow 0$, the tree-level process must be
considered together with the virtual QED correction to the resonance
process $\mu^+\mu^- \rightarrow H$ to obtain an infrared-finite
$O(\alpha)$ inclusive calculation\cite{tik}. Here, we are concerned with
production of the Higgs with an observable, relatively hard photon. In
Table I, the total cross section is shown as a function of $m_H$ for
several collider energies. At 500 GeV, luminosities of order 100
fb$^{-1}$ are needed to probe this channel. To make this statement more
precise, we investigated the principal background $\mu^+\mu^-\rightarrow
b\bar{b}\gamma$ by adapting the amplitudes for $e\bar{e}\rightarrow
\mu^+\mu^-\gamma$ \cite{eemmg}. In Table II, the background
contributions are shown for several cuts on the $b\bar{b}$ invariant
mass $m_{b\bar{b}}$. In addition to these invariant mass cuts, we
require the transverse momenta of the $b$, $\bar{b}$ and $\gamma$ to be
greater than 15 GeV, their rapidities $y$ to be less than 2.5 and the
separation $\Delta R$ between the $\gamma$ and the $b$ and the $\gamma$
and the $\bar{b}$ to be greater than 0.4. The background is compared to
the signal in Table I for Higgs boson masses of 100 GeV and 200 GeV.
This comparison shows that, while not a discovery mode for the Higgs
boson, photon-Higgs associated production can be observed with signal to
square root of background ratios $(S/\sqrt{B})$ greater than 2 when $m_H
> 100$\,GeV at a 500 GeV collider.

\begin{figure}[b!] 
\centerline{\epsfig{file=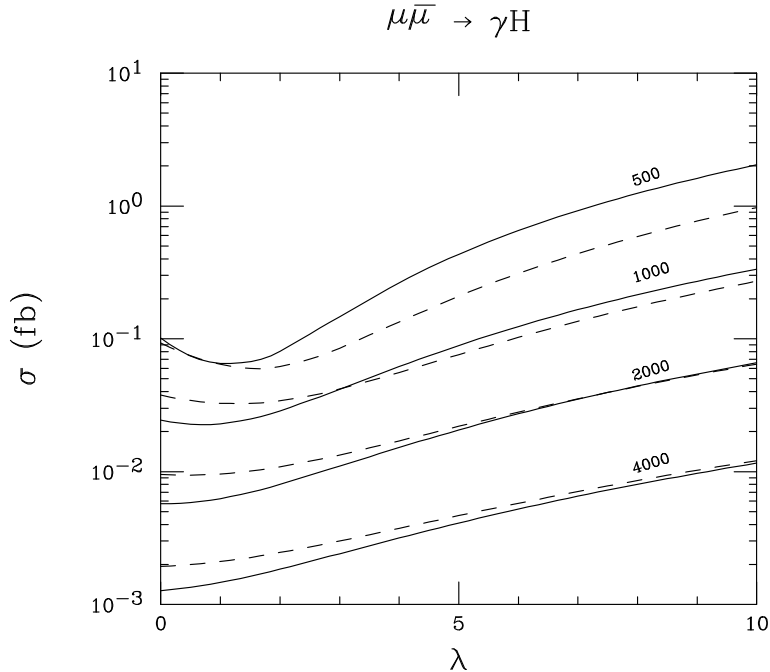,height=3.5in}}
\vspace{10pt}
\caption{The cross section for $\mu^+\mu^-\protect\rightarrow H\gamma$
obtained by scaling the Standard Model $t\bar{t}H$ coupling by a factor
$\lambda$ is shown for collider energies of 500 GeV, 1000 GeV, 2000 GeV
and 4000 GeV. In each case, the solid line is $m_H = 60$\,GeV and the
dashed line is $m_H = 250$\,GeV.}
\label{fig12}
\end{figure}

Finally, we comment on how the one-loop contribution makes it possible
to use $\mu^+\mu^-\rightarrow H\gamma$ as a probe of the Higgs-boson
coupling to $W$'s, $Z$'s and top quarks. To provide a sense of the
sensitivity of the $H\gamma$ cross section to changes in Standard Model
couplings, we have varied the $t\bar{t}H$ coupling by a factor $\lambda$
\cite{hks}. The result is shown in Fig. 5, where the characteristic
feature is the minimum in the cross section at the Standard Model value
$\lambda = 1$. For $\lambda > 1$, the cross section rises significantly.
At a 500 GeV collider, observation of $H\gamma$ production with a cross
section of order 1 fb would indicate some type of anomalous coupling.
Anomalous Higgs-gauge boson couplings have been studied\cite{gounaris}
for the case of  $e^+e^-\to H\gamma$ where enhancement of couplings such
as $HZZ$ could result in immense enhancements to the level of 10~fb.
Thus, polarization of the muon collider {\em opposite} to that required
for optimizing resonance Higgs production may be useful in highlighting
the loop-level contributions to $\mu^+\mu^-\to H \gamma$, which if
unexpectedly large could signal such anomalous couplings.

This research was supported in part by the U.S. Department of Energy
under Contract Nos. DE-FG013-93ER40757 and DE-FG02-84ER40173, and in
part by the National Science Foundation under Grant No. PHY-93-07980.

\end{document}